\documentstyle[12pt,draft]{article}

\newcommand{\beq}{\begin{equation}}
\newcommand{\eeq}{\end  {equation}}

\newcommand{\beqar}{\begin{eqnarray}}
\newcommand{\eeqar}{\end  {eqnarray}}

\newcommand{\benum}{\begin{enumerate}}
\newcommand{\eenum}{\end  {enumerate}}

\newcommand{\bfig}{\begin{figure}}
\newcommand{\efig}{\end  {figure}}

\newcommand{\btab}{\begin{table}}
\newcommand{\etab}{\end  {table}}

\newcommand{\NPA}[1]{Nucl. Phys. {\bf A{#1}}}
\newcommand{\PLB}[1]{Phys. Lett. {\bf {#1}B}}
\newcommand{\PRC}[1]{Phys. Rev. {\bf C{#1}}}

\begin{document}

\title{The Longitudinal and Transverse responses in \\
inclusive electron scattering within\\
the RPA framework.}
\vspace{17 mm}
\author{E. Bauer$^1$, A. Polls$^2$ and A. Ramos$^2$\\
$^1$Departamento de F\'{\i}sica, Facultad de Ciencias Exactas,\\
Universidad Nacional de La Plata,\\
La Plata, 1900, Argentina.\\
$^2$Departament d'Estructura i Constituents de la Mat\`eria,\\
Universitat de Barcelona,\\
Diagonal 647, E-08028, Spain.}
\maketitle
\begin{abstract}
The longitudinal and transverse nuclear responses
to inclusive electron scattering reactions are
analyzed within the Random Phase Approximation (RPA) framework.
Several residual interactions are considered
and it is shown  that the exchange terms in the RPA
make very difficult to find an effective residual
interaction capable of reproducing simultaneously the quasielastic
peak of both the longitudinal and transverse responses.
By means of a simplified model it is illustrated that the
residual interaction used in a ring approximation
must fulfill some restrictions in order to qualitatively
reproduce the full RPA results.

\end{abstract}

\vskip1cm
PACS number: 21.65, 25.30.Fj, 21.60.Jz.

Keywords: Nuclear Electron Scattering.

\newpage

In this work we present a study of the
longitudinal and transverse nuclear
response functions to inclusive electron
scattering reactions within the Random Phase
Approximation (RPA) framework in
non-relativistic nuclear matter. It is
only recently that two different procedures
to evaluate the nuclear responses in the RPA
scheme are available for a general finite range
effective interaction \cite{kn:ba96,kn:de98}. In the past,
most of the calculations have been performed in
the so-called Ring Approximation (RA).
In fact, both the RA and RPA approximations account for the
excitations of
particle-hole type that can be induced by the electromagnetic
probe but,
formally, the RA is only the direct contribution of
the RPA.
It has been common to adjust the effective
nucleon-nucleon interaction to
reproduce the nuclear response in the quasielastic peak within the
RA. We note that within the RA
different pieces of the interaction act either in the longitudinal
or in the transverse response, facilitating the extraction of the
parameters of the interaction for a certain momentum transfer.
Since the explicit evaluation of exchange terms is
important, as we will see, one could aim to apply the same procedure in
the framework of
the RPA.

To find a force that gives a reasonable account of the basic features
of both responses would be particularly interesting to
define a proper framework beyond which other
excitations,
like ground state correlations, final state interactions and meson
exchange currents, which have been shown to be important
(see \cite{kn:bo96} and Refs. therein),
are included.
In the present brief report we have explored several representative
interactions commonly used in the literature and, as we will
show, no one is able to give a simultaneous satisfactory
description of both longitudinal and transverse responses at
various momentum transfer values. The reason lies in the
fact that, as opposed to the RA, the full interaction
is present in the RPA-exchange terms
and in addition these terms are very important. Therefore
the adjustment of one channel affects the other one and this means
that reproducing both channels, longitudinal and transverse, is
a far more complex task.

On the other hand, it has been common to use a modified residual
interaction in the RA scheme trying to account
for exchange terms. Another purpose of the present work is to
show, making use of a simplified model, that while the RA
is able to keep
the qualitative trends of the
complete RPA response, it is unable to provide a quantitative
account of the exchange terms tied to the finite range part of the
force.
As it will be shown below, the qualitative agreement
can only be achieved if some simple constraints over the parameters
of the interaction
are observed.

The longitudinal and transverse pieces of the RPA
response are calculated following the
scheme of Ref. \cite{kn:ba96}, but using a HF basis according to the
prescriptions given in \cite{kn:ba96b}, where the
effect of the HF self-energy is
adjusted by means of
a set of two effective masses (one for
particles and the other one for holes) and an energy shift. Each
set is chosen to reproduce the exact HF response and
depends only on the momentum transferred by the electron.
The RPA calculation is performed for three different
interactions. The first one,
labelled $V^I$, is described in
 \cite{kn:ce94} and
contains the exchange of the mesons
$\pi$, $\rho$ and $\omega$ plus a $g'$ term.

The structure of the second one has been
widely used in the literature and consists of
contact terms plus a
($\pi+\rho$)-meson exchange interaction given by,
\beq
V^{II}(k) = \frac{f_{\pi}^2}
{\mu_{\pi}^2}
(f_0 \: + \: f'_0 \; \mbox{\boldmath $\tau \cdot \tau ' $} \: + \:
g_0  \; \mbox{\boldmath $\sigma \cdot \sigma ' $} +
g'_0 \; \mbox{\boldmath $\tau \cdot \tau ' $}
\mbox{\boldmath $\sigma \cdot \sigma ' $} \: + \:
V_{\pi} (k) \: + \: V_{\rho} (k) )
\label{eq:inter}
\eeq
with
\beq
V_{\pi} (k) \; = \;
- \, \Gamma_{\pi}^2 (k) \:
\frac{k^2}{k^2 + \mu_{\pi}^2} \ \;
\mbox{\boldmath $\sigma \cdot \widehat{k} $} \,
\mbox{\boldmath $\sigma ' \cdot \widehat{k} $}  \;
\mbox{\boldmath $\tau \cdot \tau ' $} \,
\label{eq:vpi},
\eeq
\beq
V_{\rho} (k) \; = \; - \, C_{\rho} \;
\Gamma_{\rho}^2 (k) \:
\frac{k^2}{k^2 + \mu_{\rho}^2} \ \;
(\mbox{\boldmath $\sigma \times \widehat{k} $}) \, \cdot \,
(\mbox{\boldmath $\sigma ' \times \widehat{k} $})  \;
\mbox{\boldmath $\tau \cdot \tau ' $} \,
\label{eq:vrho},
\eeq
where $\mu_{\pi}$ ($\mu_{\rho}$ )
is the pion (rho) rest mass and
$C_{\rho} = 2.3$.
For the form factor of the
$\pi NN$ ($\rho NN$ ) vertex
we have taken
\beq
\Gamma_{\pi, \rho} (k) =
\frac{\Lambda_{\pi, \rho}^2 - \mu_{\pi, \rho}^2 }
{\Lambda_{\pi, \rho}^2 + k^2} \ ,
\label{eq:ver}
\eeq
with $\Lambda_{\pi} = 1.3$ GeV and $\Lambda_{\rho} = 1.75$
GeV. The momentum transferred by the interaction is $k$ and
the static limit of the $(\pi+\rho)$-meson exchange interaction has been
taken.
The parameters $f_0,f^\prime_0,g_0,g^\prime_0$ of this force have
been left free so as to reproduce both the longitudinal and transverse
responses. However, the presence of all terms of the interaction in the
exchange contributions makes this purpose very difficult. This will
be illustrated more clearly in the simplified model discussed below.
An acceptable reproduction of both channels, especially the
longitudinal one, for momentum transfer
$q=410$ MeV/c is found with the values
$f_0=-0.1$, $f'_0=0.2$, $g_0=0$ and
$g'_0=0.5$.

Finally, we consider the parameterization of
the Bonn potential
of Ref. \cite{kn:ba96b}. This interaction,
which will be referred as $V^{III}$, is expressed as
the exchange of $\pi$, $\rho$, $\sigma$ and $\omega$ mesons.

For each interaction, the effective masses and energy shifts
neccesary to calculate the HF responses
are shown in Table I.

In Figs. 1 and 2, we present the RPA results (solid lines) for a
momentum transfer of
$q=300$ MeV/c and
$q=410$ MeV/c, respectively, and for the three interaction models.
Our results are compared
with the experimental data on $^{40}$Ca \cite{kn:wi98}.
Also we present the HF (long-dashed lines) and the RA
(short-dashed lines) responses. The latter is obtained from the ``contact"
interaction model explained below.
In all the calculations the Fermi momentum is taken to be $k_F=235$ MeV/c,
which is an appropiate value to simulate
the results in finite nuclei \cite{kn:am94}.
The HF response is in fact qualitatively similar to the free Lindhard
function but it is hardened by the presence of the single particle
spectrum, which increases the energy of particle-hole excitations.

The behavior of the RPA responses in the longitudinal and transverse
channels
depends on the interaction considered.
For the $V^{I}$ and  $V^{II}$ interactions,
when the position of the RPA peak
for one channel is moved towards higher
energies with respect to the peak in the HF response, the peak in
the other channel moves in
the opposite direction.
This particular feature is a consequence of
the exchange terms in the RPA.
Only the $V^{III}$ interaction gives a hardening of the response
in both channels.
We also observe that the interaction
$V^{II}$, whose parameters have been adjusted to account
for the response at $q=410$ MeV/c, produces a response
at $q=300$ MeV/c in poor agreement with the experimental
results.  This means that the interaction should have
a richer momentum dependence than just the one provided
by $\pi$- and $\rho$- meson exchange.

In order to clarify this behavior, let us
consider the simpler case of an
interaction:
\beqar
V_C & = & \frac{f_{\pi}^2}
{\mu_{\pi}^2}
(f \: + \: f' \; \mbox{\boldmath $\tau \cdot \tau ' $} \: + \:
g  \; \mbox{\boldmath $\sigma \cdot \sigma ' $} +
g' \; \mbox{\boldmath $\tau \cdot \tau ' $}
\mbox{\boldmath $\sigma \cdot \sigma ' $} +
\nonumber \\
& & \: + \: h \; \mbox{\boldmath $\sigma \cdot \widehat{q} $} \,
\mbox{\boldmath $\sigma ' \cdot \widehat{q} $}  \;
\: + \: h' \; \mbox{\boldmath $\sigma \cdot \widehat{q} $} \,
\mbox{\boldmath $\sigma ' \cdot \widehat{q} $}  \;
\mbox{\boldmath $\tau \cdot \tau ' $} \,)
\label{eq:intcon}
\eeqar
where $f$, $f'$, $g$, $g'$, $h$ and $h'$ are all constants.
We have
preferred to use the spin longitudinal terms $h$ and $h'$
instead of the tensor terms (proportional to $S_{12}$).
This is because, at variance with $S_{12}$, the spin
longitudinal terms do not interfere with the vector terms
$g$ and $g'$ in the RA.

This interaction is now used to evaluate direct
and exchange terms in the RPA. For such an
interaction it is possible to account
for exchange terms by a re-definition of these
constants as follows
\beqar
f_{ant} & = & f \, - \, (f + 3 f' + 3 g + 9 g' + ~~h + 3 h')/4
\nonumber \\
f'_{ant} & = & f' \, - \, (f - ~~f' + 3 g - 3 g'+ ~~h - ~h' )/4
\nonumber \\
g_{ant} & = & g \, - \, (f + 3 f' - ~~g - 3 g'- ~~h - 3 h')/4
\nonumber \\
g'_{ant} & = & g' \, - \, (f - ~~f' - ~~g + ~g' - ~h + ~h')/4
\nonumber \\
h_{ant} & = & h \, - \, (~h + 3 h')/2
\nonumber \\
h'_{ant} & = & h' \, - \, (~h - ~h')/2
\label{eq:cont}.
\eeqar

The first term in the {\it r.h.s.} of each equation is the direct
contribution, while the terms between parenthesis
come from the action of the exchange operator over the interaction.
It is important to keep in mind that solving the RA with this new
set of parameters is equivalent to solving the
RPA with the original contact interaction of Eq. (\ref{eq:intcon}).
At variance with the direct case, only three of
the parameters $f_{ant}$-$h'_{ant}$,
are independent.
They are related through the following relations:
\beqar
g'_{ant} & = & - \, (f_{ant} \, + \, h'_{ant} ) /3
\nonumber \\
g_{ant} & = & - (2 \, f_{ant} \, - \, h'_{ant} ) /3 \, - \, f'_{ant}
\nonumber \\
h_{ant} & = & - \, h'_{ant}
\label{eq:con2},
\eeqar
which are obtained by simply solving the system (\ref{eq:cont}).
Note that the spin longitudinal terms are usually assumed to be
proportional to the momentum transfer and they cancel in the Landau limit.
Formally, Eq. (\ref{eq:intcon}) with
$h=h'=0$, can be viewed as the zeroth order in the Legendre
expansion of the Landau-Migdal interaction. In this sense,
(\ref{eq:con2}) is a particular case of the more general sum rules
results of Ref. \cite{kn:fr79}.

Within the RA, $f_{ant}$ and $f'_{ant}$ govern with the same weight
the longitudinal response, while
for the transverse one only $g'_{ant}$ is relevant
($g_{ant}$ gives also a contribution, but of
the order $\mu_s^2 / \mu_v^2 \approx 0.035$
with respect to that of $g'_{ant}$).

Now we turn
back to the analysis of the RPA
response with a general finite range interaction. For direct
RPA terms, the external momentum fixes the momentum
of the  particle-hole interaction. That means that, for direct
RPA terms, the interaction behaves as a contact one for
each momentum transfer. In Table II we have extracted the
``contact" terms of the interactions $V^{I-III}$ for $q=300$ MeV/c
and $q=410$ MeV/c.
Working as if these parameters belonged to a contact interaction we have
obtained, using Eqs. (\ref{eq:cont}), the corresponding values
$f_{ant}$-$h'_{ant}$, also shown in Table II.
We have then performed a RA calculation with these values and the
results are given by the short-dashed lines in Figs.~1-2.
This simple model illustrates
that if we represent a general interaction by a contact one by ignoring
the momentum dependence of the finite range terms, the corresponding
RPA (which is in fact a RA with the parameters redefined according to
Eqs.~(\ref{eq:cont})) gives a qualitative
agreement with the result obtained using
the complete interaction, which is displayed by the solid lines in
Figs.~1-2.
The position and shape of the quasielastic peak are qualitatively similar,
in both longitudinal and
transverse channels.
Therefore, the momentum
dependence of the finite range
force in the exchange terms is not strong enough to modify the trends
on the longitudinal and transverse responses induced by the simplified
contact interaction.
This, in turn, implies that the coefficients of the different terms of
any effective interaction that
might be used to account effectively for exchange terms in RA should obey to
a high degree the restrictions imposed by the relations (\ref{eq:con2}).

We can summarize our results by saying that the momentum dependence
of the interaction in the exchange terms gives rise to quantitative
differences between the RPA and the RA responses, the later one being
calculated with the antisymmetrized contact version of the original
force. This makes advisable to use
the RPA with the complete interaction before attempting
more complicated studies of other types of correlations.
Correlations beyond RPA are certainly necessary since,
to the best of our knowledge, no work is able to
reproduce the quasi-elastic response at any momentum transfer
for both longitudinal and transverse channels, using the same
interaction for medium and
heavy nuclei. We have unsuccessfully attempted to find one
interaction that at least reproduces the position of the peaks for
both channels within the RPA framework.

Another observation of the present work is that if one
still wants to account for exchange terms within a simpler RA approach one
should make sure that the effective interaction is such that the
relations (\ref{eq:con2}) are preserved to a high degree,
since only in this case the RPA and RA are in qualitative agreement.
The implementation of these relations is as simple as the
RA itself.
We consider significant this partial saving
of the RA because of the extreme simplicity of the ring
propagator in nuclear matter, which contrasts with the
difficult evaluation of the full RPA.

It is also important to note that,
while it is possible
to find an interaction that adjusts the longitudinal and transverse
channels in a RA scheme, it will not obey the relations
(\ref{eq:con2}) and, therefore, the corresponding original direct
interaction cannot
be recovered.
Such type of interactions cannot be blindly used in
calculations of other types of many-body correlations
that explicitly require the knowledge of the original interaction.
In particular, one should avoid replacing the effective
interaction by dressing it with the ring propagator if relations
(\ref{eq:con2}) are ignored.

\newpage

\vfill
\eject
\section*{Figure Captions}
\begin{description}

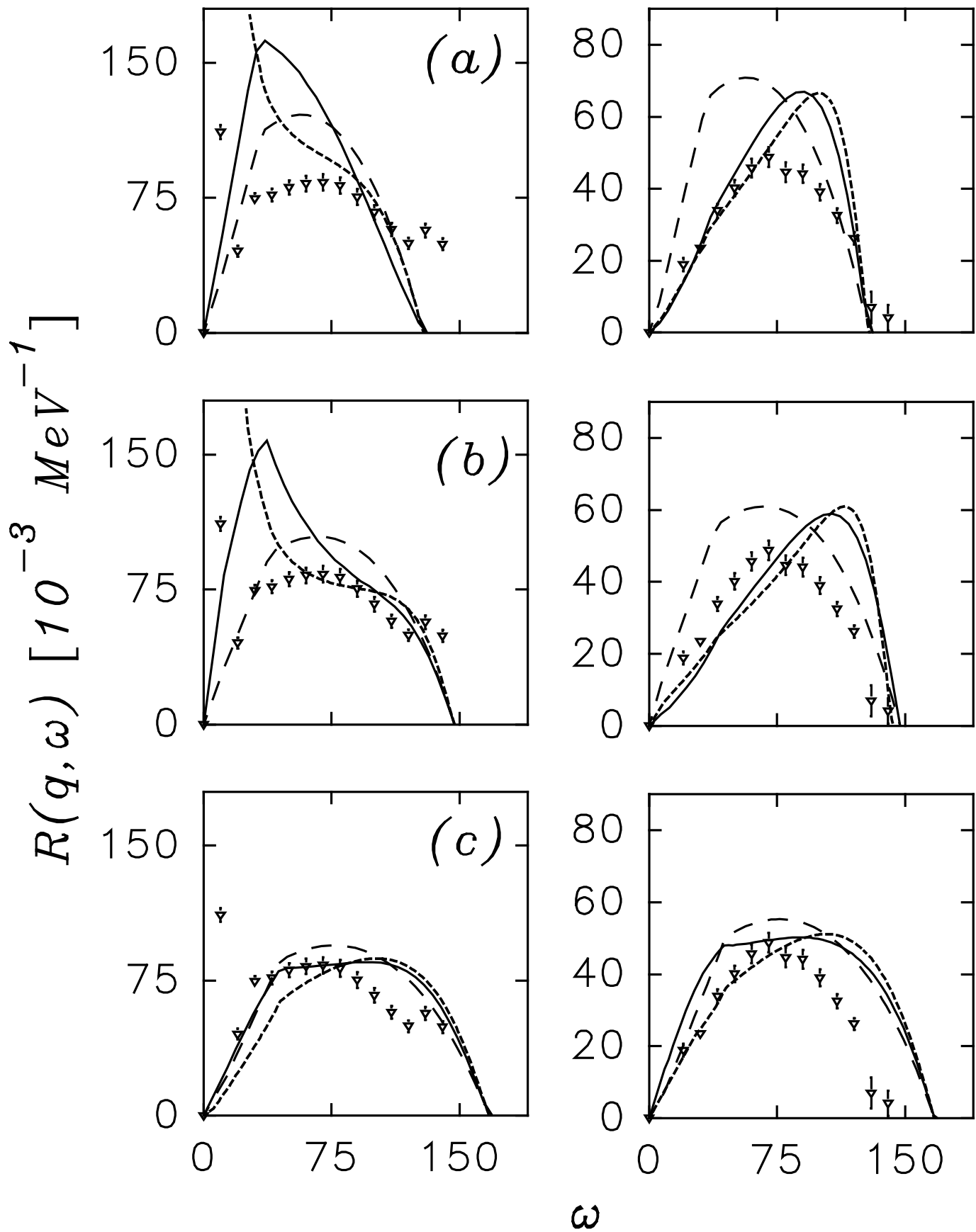
\item [Figure 1:]
Response function for $^{40}$Ca at momentum transfer
$q=300$ MeV/c.  The panels on the {\it l.h.s.}
show the longitudinal response and those on
the {\it r.h.s.} show the transverse one.
Results are shown for the HF (long dashed
line), the RA (short-dashed line) and the
RPA responses (solid line).
Results in rows  (a), (b) and (c) are
obtained with the $V^I$, $V^{II}$ and $V^{III}$ interactions, respectively.
The experimental data was taken
from \cite{kn:wi98}.

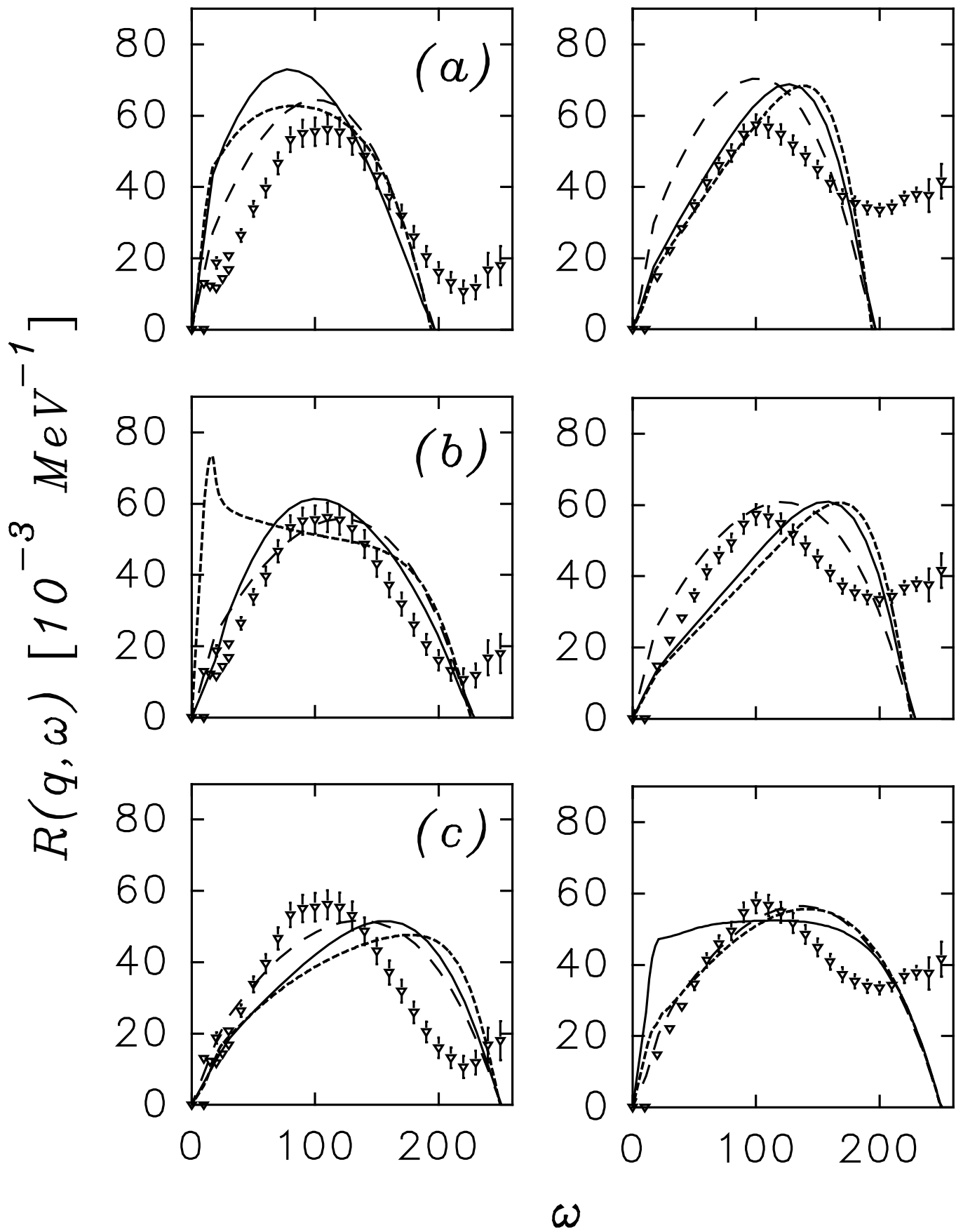
\item [Figure 2:]
The same as Fig. 1, but
for a momentum transfer
$q=410$ MeV/c.

\end{description}
\vfill
\eject

\newpage

\begin{quote}

\centering Table I.

\vspace{7 mm}

\begin{tabular}{cccc}         \hline\hline
$q=300$ MeV/c & & &  \\ \hline
int. & $m^*/m $ &   & $\Delta \omega$ [MeV]   \\
\cline{2-3}
& particles & holes & \\ \hline
$V^{I}$     & 0.97 & 0.85  & 6.~ \\
$V^{II}$    & 0.83 & 0.67  & 7.~ \\
$V^{III}$   & 0.75 & 0.65  & 9.~ \\ \hline\hline
$q=410$ MeV/c & & &  \\ \hline
 int. & $m^*/m $ &   & $\Delta  \omega$  [MeV]   \\
\cline{2-3}
& particles & holes & \\ \hline
$V^{I}$     & 1.0~ & 0.85  & 7.~ \\
$V^{II}$    & 0.85 & 0.67  & 9.~ \\
$V^{III}$   & 0.78 & 0.65  & 11. \\ \hline\hline
\end{tabular}

\end{quote}

\vspace{15 mm}

Table I: Effective masses and energy shifts for the different
residual interactions described in the text and for two values of the
momentum transfer.
The energy shift is denoted by $\Delta \omega $.

\newpage

\begin{quote}

\centering Table II.

\vspace{7 mm}

\begin{tabular}{ccccccc}     \hline\hline
$q=300$ MeV/c \\ \hline
int.&$~~f~~$&$ ~~f'~~$&$~~g~~$&$~~g'~~$& $~~h~~$&$~~h'~~$ \\ \hline
$V^{I}$     &~0.000  &~0.000  &  -0.094 & ~0.375 & ~0.094 & -0.465 \\
$V^{II}$    &-0.100  & ~0.200 &  ~0.000 & ~0.351 & ~0.000 & -0.578 \\
$V^{III}$   &-0.630  & ~0.085 &  -0.085 & -0.163 & ~0.085 & -0.570 \\
 \hline
       & $f_{ant}$ &$f'_{ant}$  &$g_{ant}$ & $g'_{ant}$
&$h_{ant}$ & $h'_{ant}$ \\ \hline
$V^{I}$     &-0.448  & ~0.212 &  -0.162 & ~0.397 & ~0.744 & -0.744 \\
$V^{II}$    &-0.581  & ~0.394 &  -0.295 & ~0.483 & ~0.868 & -0.868 \\
$V^{III}$   &~0.300  & ~0.042 &  -0.541 & ~0.199 & ~0.898 & -0.898 \\
 \hline\hline
$q=410$ MeV/c \\ \hline
int.&$~~f~~$&$ ~~f'~~$&$~~g~~$&$~~g'~~$& $~~h~~$&$~~h'~~$ \\ \hline
$V^{I}$     &~0.000 & ~0.000 & -0.148 & ~0.282 & ~0.148 & -0.386 \\
$V^{II}$    &-0.100 & ~0.200 & ~0.000 & ~0.267 & ~0.000 & -0.493 \\
$V^{III}$   &-0.442 & ~0.041 & -0.134 & -0.251 & ~0.134 & -0.482 \\
 \hline
       & $f_{ant}$ &$f'_{ant}$  &$g_{ant}$ & $g'_{ant}$
&$h_{ant}$ & $h'_{ant}$ \\ \hline
$V^{I}$     &-0.270 & ~0.189 & -0.226 & ~0.308 & ~0.653 & -0.653 \\
$V^{II}$    &-0.455 & ~0.352 & -0.295 & ~0.398 & ~0.739 & -0.739 \\
$V^{III}$   &~0.630 & -0.080 & -0.603 & ~0.053 & ~0.789 & -0.789 \\
 \hline\hline
\end{tabular}

\end{quote}

\vspace{15 mm}

Table II: For a fixed momentum the interaction for direct RPA terms
is a constant. Columns $f$-$h'$ are the values of these constants
for the interactions $V^{I-III}$ described in the text. Columns
$f_{ant}$-$h'_{ant}$ result
from applying Eqs.~(\ref{eq:cont}) using the
values of columns $f$-$h'$ as direct contributions.


\newpage

\begin{thebibliography}{99}

\bibitem {kn:ba96}
E. Bauer, A. Ramos and A. Polls, \PRC{54} (1996) 2959.

\bibitem {kn:de98}
A. De Pace, \NPA{635} (1998) 163.

\bibitem {kn:bo96}
S. Boffi, C. Giusti, F. D. Pacati and M. Radici;
Electromagnetic Response of Atomic Nuclei, Clarendom Press,
Oxford 1996.

\bibitem {kn:ba96b}
M. B. Barbaro, A. De Pace, T. W. Donnelly and A. Molinari,
\NPA{596} (1996) 553.

\bibitem {kn:ce94}
R. Cenni and P. Saracco,  \PRC{50} (1994) 1851.

\bibitem {kn:wi98}
C. F. Williamson {\it et al.}, \PRC{56} (1998) 3152.

\bibitem {kn:am94}
J. E. Amaro, A. M. Lallena and G. C\'o,
Int. J. Mod. Phys. {\bf E3} (1994) 735.

\bibitem {kn:fr79}
B. L. Friman and A. K. Dhar, \PLB{85} (1979) 1.

\bibitem {kn:ba99}
E. Bauer and A. Lallena, \PRC{59} (1999) 2603.

\end{thebibliography}
\end{document}